# The effect of alternating current on the current states of a quantum interferometer shunted by a superconducting inductance

S. I. Link, V. P. Koverya, A. V. Krevsun,[a)] and S. I. Bondarenko

*B.I. Verkin Institute for Low Temperature Physics and Engineering of the National Academy of Sciences of Ukraine 47 Nauky Ave, Kharkiv, 61103, Ukraine*

The patterns of reversible changes in the critical current and discrete current states of a structure in the form of a superconducting quantum interferometer shunted by superconducting inductance, as a result of passing an alternating transport current through the structure and applying an external alternating magnetic field simultaneously with a direct transport current, was established. A new type of discrete stationary state was discovered during the transition of the interferometer to the resistive state caused by the combined action of direct and alternating transport currents.

## 1. Introduction

This paper is dedicated to the study of current characteristics of a superconducting quantum interference device (SQUID) shunted by a larger superconducting inductance (SI) compared with that of the SQUID (see Fig. 1).

Shunting of a SQUID with a SI was proposed by us several years ago.[1–4]

As shown in Refs. 2 and 3, the use of this structure makes it possible to simultaneously measure the energy gap and the relaxation time of the superconducting state in superconductors. In this case, the measurement accuracy relevant to these basic parameters of superconductors can be significantly improved over that of previously known methods. A mathematical description of the processes taking place in the SQUID-SI structure is currently under development. Various model representations of these processes are analyzed in this work.

Other authors have previously proposed and extensively investigated a shunted SQUID using ohmic resistances,[5] electric capacities[6] and inductances of non-superconducting metals.[7] The creation of such structures made it possible to manage various electrical characteristics of a SQUID with the aim of expanding the scope of their application in science and technology.

The purpose of this work was to experimentally study new characteristics of the SQUID-SI structure that yield further information about its current states. In contrast to the basic characteristics of the structure, each of the new characteristics binds together three electrical parameters rather than two. Specifically, dependency $I_1(I)$ acquires a new form with an additional impact on the structure, particularly by imposing field $H$ on the circuit with a SI, or by adding alternating transport current $I^\sim$. A total of four new dependencies were studied (see sections 2 and 3 below), which made it possible to establish current quantum states of the structure under the integrated action of direct and alternating transport currents, as well as static and alternating magnetic fields.

## 2. Regarding the Theory of Basic Characteristics of the SQUID-SI Structure

An important work that served as the starting point for the creation and study of the SQUID-SI structure was a fundamental article by Silver and Zimmerman, published 50 years ago.[8] The article investigates current quantum states of the simplest SQUID with one superconducting point contact [Fig. 2(a)].

Imposition of external magnetic field $H_e$ on the SQUID circuit generates circulating current $i$ inside, which, with the increase of $H_e$, reaches $i_c$.

$$L_0 i_c/(\Phi_0/2) = 1, \qquad (1)$$

Subject to the following condition where $L_0$, $i_c$, $\Phi_0$ are the inductance, the critical current of the SQUID circuit, and the magnetic flux quantum, respectively, there is a periodic dependence of current $i$ on the external magnetic flux $\Phi_e$ [Fig. 2(b)] with discrete transitions of current $i$ at semi-quantum values of the external magnetic field flux with period $\Delta\Phi_e = \Phi_0$. In this case, the magnetic flux $\Phi$ in the SQUID circuit is also changed discretely by the quanta of the flux $\Phi_0$.[8] During the circulating current step in a SQUID, it momentarily attains voltage and resistivity $R$ [Fig. 2(c)]. Similar dependencies $i(\Phi)$ and $\Phi(\Phi_e)$ also take place in the case where magnetic flux $\Phi_e$ is created by direct transport current $I$ passing through this SQUID [Fig. 2(a)].[8]

The main electrical characteristics of the SQUID-SI structure include: the dependence of current ($I_1$) through a SI on the magnitude of direct transport current ($I$) through the SQUID[1,2] and the dependence of current $I_1$ on an external constant magnetic field ($H$) through the superconducting circuit with a SI at zero transport current ($I=0$). A typical view of these dependencies is shown in Fig. 3.[4]

To analyze the processes in the SQUID-SI structure, it is first necessary to consider the peculiarities of using a two-contact SQUID, through which direct transport current $I$ is passed [Fig. 2(e)]. In order for this current to create magnetic flux $\Phi_e$ through the contour of this SQUID, it should be asymmetric in the magnitude of the critical contact currents. It should be noted that the appearance of the resulting magnetic flux $\Phi_e$ in a superconducting circuit with a relatively large inductance ($10^{-8}$ H) with two film bridge contacts having differing critical currents, one in each of its branches, was recently confirmed by our direct measurements of the magnetic field on the surface of this circuit.[9] It was also shown that the magnetic flux in this circuit appears rather smoothly, i.e. without a jump, and corresponds to the critical current of a weaker bridge.

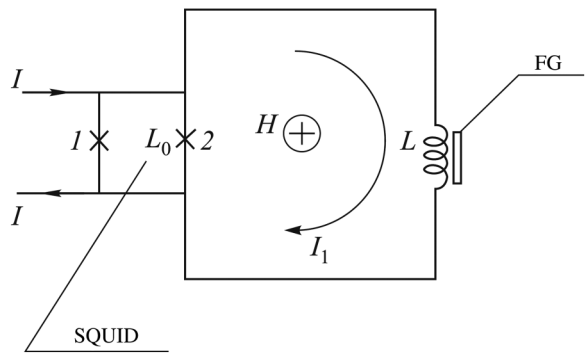

Fig. 1. The SQUID-SI structure; *1, 2* are Josephson contacts, $I$ is an incoming transport current, $I_1$ is a current in a large superconducting inductance $L$, FG is a fluxgate magnetometer sensor.

The SQUID-SI structure shown in Fig. 1 is in fact a large superconducting circuit, with one of its branches being an asymmetric two-contact SQUID with an extremely small inductance ($L_0 \sim 10^{-13}$ H),[2,3] and the other being its part with a significantly greater inductance ($L = 10^{-6}$ H).

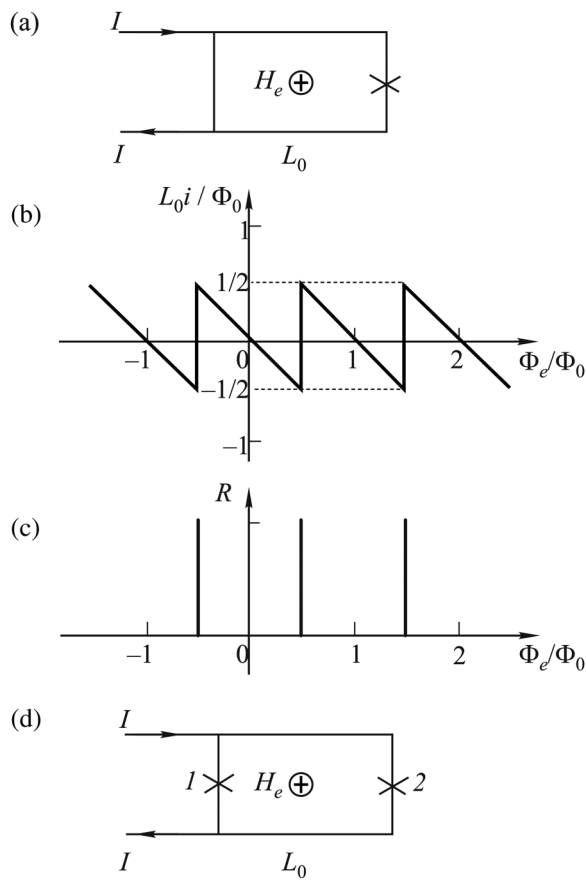

Fig. 2. Diagram of a single-contact SQUID (a) and its characteristics as a function of the external magnetic flux $\Phi_e$ of field $H_e$ (provided that $L_0 i_c/(\Phi_0/2) = 1$, see below) for the rated values: circulating current ($i$) of the interferometer[8] (b) and the impulsive arising resistivity ($R$) of SQUID (c), the diagram of a two-contact SQUID (d).

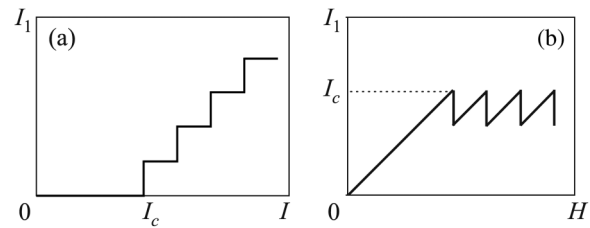

Fig. 3. Schematic view of typical basic characteristics of the SQUID-SI structure with critical current $I_c$: dependencies of current $I_1$ in a branch with inductivity $L$ (Fig. 1) on transport current $I$ (a) and on external magnetic field $H$ (b).

Let us consider the current processes in this structure. First, we will attempt to explain the appearance of the first of the primary dependencies, namely $I_1(I)$, shown in Fig. 3(a).

In most of the studied samples of this type of structure, the critical current corresponding to the first jump in the dependence $I_1(I)$, was significantly greater than the step length along the direction of current $I$. At the same time, the length and height of the steps are equal to each other, and their period on current $I$ corresponds to that of interference voltage changes on the circuit-isolated SQUID (in the form of a pressure point contact (PPC)) in the resistive state, if the current is passed through one of the microwires constituting the SQUID. Such interference dependencies can also be obtained by applying an external magnetic field to an isolated SQUID. This proves that, at current $I$ larger than critical current ($I_c$), periodic steps in the dependency $I_1(I)$ are the result of periodic changes in the quantum current states of an asymmetric SQUID under the action of a magnetic flux created inside by current $I$. Now, consider the state of the SQUID at $I < I_c$. In a certain significant range of the transport current growth from zero, such current does not create magnetic flux $\Phi$ inside, subject to the Laue law on the current distribution in a superconducting doubly connected circuit of the SQUID. The first to reach a critical state is one of two microcontacts of the SQUID (Fig. 1), having a lower critical current (denoted as $I_{c1}$). In this case, the SQUID as a whole remains in the superconducting state. A further increase in current $I$ causes the appearance of flux $\Phi$ in the SQUID, since the transport current through the first contact does not change, but only begins to flow through the second contact with larger critical current $I_{c2}$. As a result, an additional diamagnetic circulating current $i$ is generated in the SQUID circuit, similar to that shown in Fig. 2(b). If this addition ($\Delta i$) has a value corresponding to the ratio

$$L_0 \Delta i / (\Phi_0/2) = L_0(I_{c2} - I_{c1})/(\Phi_0/2) = 1, \qquad (2)$$

then, we suppose that the quantum of flow $\Phi_0$ will abruptly enter into the SQUID circuit, just as in the case of a single-contact SQUID according to relation (1). As can be seen from (2), when a two-contact asymmetric SQUID is used as part of the SQUID-SI structure, the appearance of a critical quantum state in it must be determined by the difference of the critical currents in the SQUID microcontacts, rather than by their absolute values. As soon as the flux quantum enters the SQUID, as previously shown in Ref. 2, there is a voltage pulse $V_i = \Phi_0/\delta t$; $\delta t$ is the pulse duration, $\delta t \approx L/R$, $R$ is the SQUID resistivity at the time of the impulse. The appearance of resistivity in the SQUID leads to a jump-like switching of a portion of transport current $\Delta I = \Delta i = I_{c2} - I_{c1} = \Phi_0/2L_0$ into the SI circuit [Fig. 3(a)]. Current $I_1$ in the SQUID increases by $\Delta I_1 = \Phi_0/2L_0$. Superconductivity is then restored in the SQUID; however, the current inside becomes zero, since the returned transport current, equal to $\Delta I$, is compensated by the oppositely directed current $\Delta I_1$, which is equal in value and frozen in the large circuit of the SQUID-SI structure. With further amplification of transport current ($I > \Phi_0/2L_0$) in the SQUID, a diamagnetic current appears, while frozen current $\Delta I_1$ is conserved in the SI. The frozen current $I_1$ is conserved with further amplification of current $I$ until a new jump in the circulating current is reached; this is shown by a 'plateau' in the dependence $I_1(I)$. A further increase in transport current leads to a periodic repetition of the above described process and to the periodic formation of the further 'steps' in this dependency.

Now consider the current processes in the formation of dependency $I_1(H)$. Here, we assume that an asymmetric SQUID in the SQUID-SI structure has the same properties as those given for the SQUID with a transport current. The increase of the external magnetic field $H$ creates the magnetic flux $\Phi_1$ in the large structure circuit (with a SI), exciting diamagnetic current $I_1$ in the circuit:

$$I_1 = \Phi_1/L = (\mu_0 HS)/L, \quad (3)$$

where $\mu_0$, $S$, $L$ are the magnetic permeability of air, the large circuit area of the SQUID-SI structure, and the inductance of this circuit, respectively. This current, in turn, excites the magnetic flux $\Phi_e \equiv \Phi_2$ in the SQUID circuit. The $\Phi_2$ flux induces the generation of diamagnetic circulating current $i = \Phi_2/L_0$ in the SQUID circuit. Subject to relation (2), diamagnetic current $i$ first increases in proportion to the growth of $H$, until reaching relation $\Delta i = (I_{c2} - I_{c1}) = \Phi_0/(2L_0)$. At this moment, current $i$ in the SQUID abruptly changes its direction from diamagnetic to paramagnetic. At this moment, the magnetic flux in the SQUID is $\Phi_0/2$. Simultaneously with surge current $i$ in the SQUID, a voltage pulse should be generated on it and a short-term resistivity should appear in the SQUID-SI circuit. The $\Phi_0$ flux quantum enters into the large circuit of the structure. Current $I_1$ decreases by $\Delta I_1 = (I_{c2} - I_{c1})$. The structure then returns to the superconducting state. The magnetic field $H$ acting at that time induces a diamagnetic current in the SQUID, which compensates for the above-mentioned paramagnetic current. As a result of changing the current flow by half a quantum, there is no current passage through the SQUID.

A further increase in the field again excites the diamagnetic currents in the large circuit and in the SQUID, i.e. the current process repeats itself. As a result, there appears a periodic repetition of sharp ('sawtooth-like') drops of the diamagnetic current in the large circuit in accordance with the current model of current processes. The dependency in Fig. 3(b) corresponds to the above current changes in the structure.

It also follows from the above that the modulation depth of current $I_1$ in dependency $I_1(H)$ must be equal to the size of the steps in the dependency $I_1(I)$.

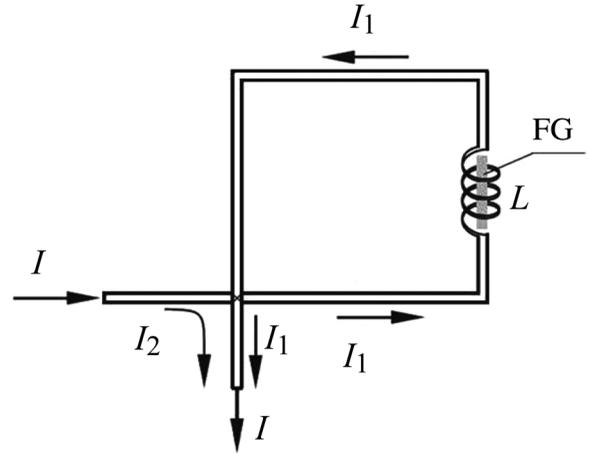

Fig. 4. Experimental structure of the SQUID-SI: $I_2$ is the direction of transport current $I$ until it reaches its critical value in the pressure contact, and $I_1$ is the direction of the transport current after it reaches its critical value in the contact.

## 3. Experimental set up

In this work, the SQUID-SI structure is a cylindrical coil with inductance $L \approx 10^{-6}$ H, made of a superconducting niobium microwire with a diameter of 70 μm, with the coil ends compressed at their intersection to form a pressure point contact according to Fig. 4. Inside the coil, there is a fluxgate magnetometer sensor that measures the magnetic field in the coil. The diagram for current measurement of this structure is shown in Fig. 5.

The structure of the SQUID-SI is formed in a manner similar to that described in our earlier works.[1–3] A PPC is one ($k = 1$) or several ($k > 1$) microcontacts in the place where the niobium microwires are pressed. The exact number depends on the surface structure of the microwire and the strength of its compression.[8] In the case where there are two microcontacts ($k = 2$), such a PPC forms a two-contact DC SQUID.[5] For the purpose of the study, such PPCs were selected, whose periodic dependence of the voltage on an external magnetic field, when exposed to liquid helium, was typical of a two-contact SQUID.[5] A fluxgate magnetometer sensor (FG) with a sensitivity of $10^{-5}$ was placed inside the cylindrical superconducting coil which is a constituent part of the SI (see Fig. 5).

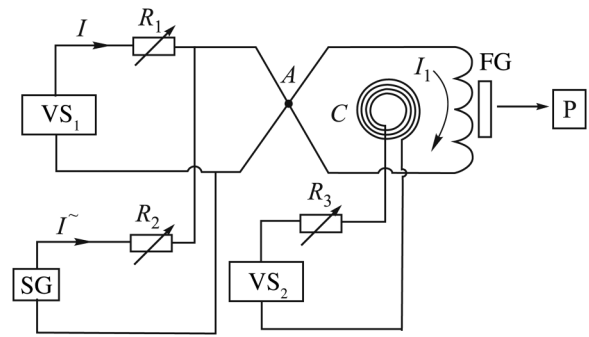

Fig. 5. Electrical circuit diagram: $VS_1$ is a voltage source for generating current $I$; $VS_2$ is a voltage source for creating the current in a flat coil $C$ generating the magnetic field $H$; SG is an audio-frequency generator for generating alternating current $\tilde{I}$ in the form of a transport current or as a current through the flat coil $C$; $R$ is a recorder of the magnetic strength measured with the help of the fluxgate FG, and $A$ is a niobium-niobium pressure point contact formed at the intersection of niobium microwires.

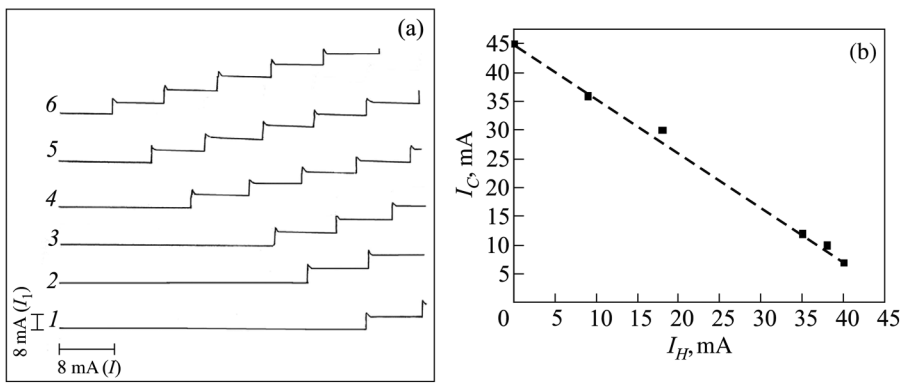

Fig. 6. Dependencies of current $I_1$ in inductance $L$ on transport current $I$ at different values of current $I_H$, mA: 0 (*1*), 9 (*2*), 18 (*3*), 35 (*4*), 38 (*5*), 40 (*6*), excited in the circuit of the SQUID-SI structure by the external magnetic field $H$ (a), dependency of the critical current of the SQUID on current $I_H$ (b).

Magnetic field of this current. To determine the current arising in this coil, pre-calibration of the magnetometer was carried out by passing a known current through this coil and taking the magnetometer readings. Transport direct ($I$) and alternating ($\tilde{I}$) currents were supplied to the SQUID according to the diagram shown in Fig. 5. The adjustment range for DC and AC was 0–1 A and 0–50 mA, respectively. The AC frequency was set in the range of 20–75 Hz, using a low-frequency sound generator. The magnetic field ($H$) was created by passing a DC through a flat copper coil with an outer diameter of 8 mm. The coil was placed in the plane of the SI circuit, covering an area of approximately 1 cm$^2$ between the PPC and the FG sensor. The magnetic axes of the FG sensor and the coil were mutually perpendicular, which excluded the direct effect of the coil field on the FG sensor. The ratio between the magnetic field $H$ generated by the flat coil and the current $I_H$ excited by this field in the SQUID was measured, taking into account the dimensions of the coil and the SQUID-SI structure, as well as the magnetic coupling between them. This made it possible to express the effect of the field $H$ on the SQUID-SI structure using the magnitude of current $I_H$, when analyzing the characteristics of the structure (see below). The SQUID-SI structure was in liquid helium at normal pressure ($T = 4.2$ K). To protect the refrigerator with the structure against external electromagnetic interference in the laboratory, it was placed in a two-layer permalloy screen. The current characteristics of the structure were recorded using the two-coordinate electromechanical recorder H-307.

## 4. Experimental results and discussion

First, consider dependency $I_1(I)$ at different values of $H$ (Fig. 6).

It is seen that the increase in $I_H$ decreases the critical current in the SQUID. The critical current was determined by the appearance of the first step of current $I_1$. Based on the abovementioned concepts of the processes in this structure, the critical state of the structure in the form of a jump of current $I_1$ occurs at current I, when the magnetic flux $\Phi$ in the SQUID circuit, created by this current, reaches the half-integer value of the $\Phi_0$ flux quantum. The formation of this flow $\Phi$ is possible, as stated in Section 2, if the SQUID is asymmetric. Figure 6(b) shows the experimental dependence of the critical current of the SQUID on the magnitude of $I_H$. As it turned out for the particular structure studied, its critical current decreases by one current step with an increase in the external magnetic field $H$ by 0.1 Oe. With the chosen directions of the field and current $I$, a linearly decreasing dependence of the critical current of the SQUID-SI structure on $I_H$ is implemented. This dependence can be explained by the summation of current $I_H$ excited by the field in the large SQUID-SI circuit, and transport current $I$. In this case, the field and transport currents have the same direction, while the field current 'withdraws' part of the superconducting current through the SQUID. As a result, the critical state of the SQUID is achieved at a lower value of transport current $I$. Specifically, the sum $(I + I_H)$ remains unchanged:

$$I + I_H = I_{c0}, \qquad (4)$$

where $I_{c0}$ is the value of the critical current at $H = 0$.

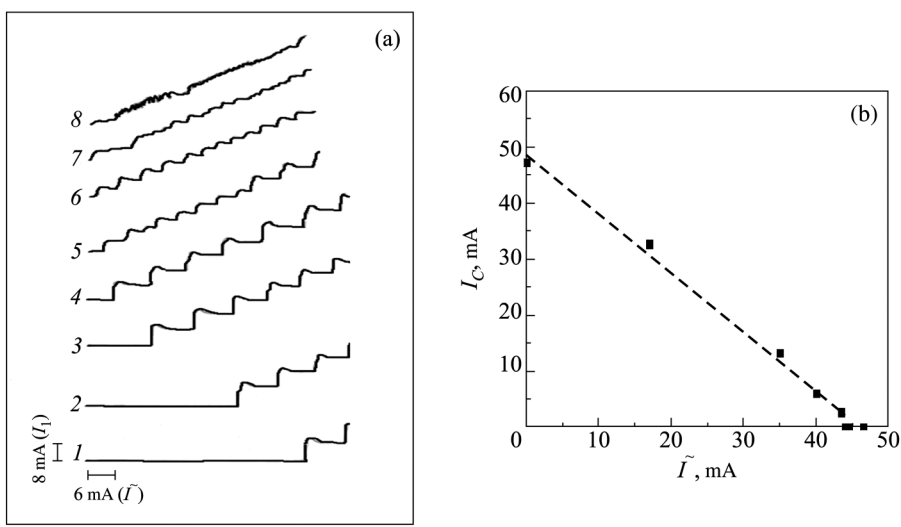

Fig. 7. Dependencies of current $I_1$ in inductance $L$ on transport DC at different amplitudes of alternating transport current $\tilde{I}$, mA: 0 (*1*), 17 (*2*), 35 (*3*), 40 (*4*), 43.5 (*5*), 44 (*6*), 44.5 (*7*), 46.5 (*8*) (a); dependence of critical DC $I_c$ of the SQUID on the amplitude of AC $\tilde{I}$ with a frequency of 75 Hz (b); $I_{c0} \approx 43$ mA.

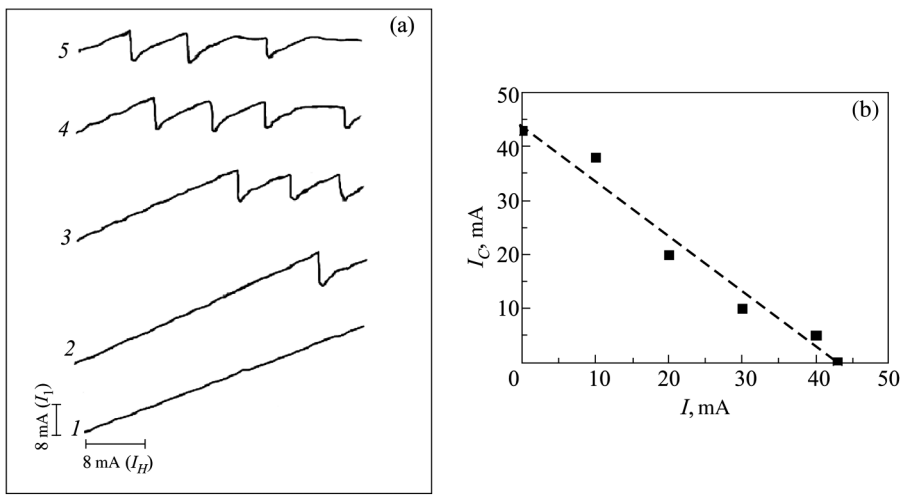

Fig. 8. Dependencies of current $I_1$ in inductance $L$ on current $I_H$ excited by the external magnetic field $H$ in the circuit of the SQUID-SI structure at different values of transport DC $I$, mA: 0 (*1*), 10 (*2*), 20 (*3*), 30 (*4*), 40 (*5*) (a); Dependence of critical current $I_c$ of the structure on transport DC $I$ (b).

Dependency $I_c(I_H)$ in Fig. 6(b) can thus be represented by the following relation

$$I_c = I_{c0}(1 - I_H/I_{c0}). \qquad (5)$$

Now, consider dependency $I_1(I)$ at different values of alternating transport current $I^\sim$ (Fig. 7).

The amplification of $I^\sim$ first leads to a linear decrease in the critical current of the SQUID (similar to the dependencies $I_1(I)$ with increasing $I_H$), as can be seen in Fig. 6(b). This section of the dependency can be described as

$$I_c = I_{c0}(1 - I^\sim/I_{c0}). \qquad (6)$$

This dependence can be explained by the fact that direct current $I$ is combined with an AC half-wave of the same direction and generates the first step of current $I_1$ at ever lower values of DC $I_c$. The opposite half-wave cannot affect the value of the critical current because it is subtracted from the value of current $I$. Starting from the value $I^\sim \approx I_{c0}$, dependencies $I_1(I)$ begin to drastically differ from those at smaller values of $I^\sim$. Current $I_1$ appears in steps of smaller sizes and a smaller period in between, until they completely disappear with a relatively small excess of ($\delta$) amplitudes of alternating current $I^\sim$ over the initial critical DC SQUID $I_{c0}$ ($I^\sim \approx I_{c0} + \delta$, where $\delta$ is the step length at $I^\sim < I_{c0}$). This effect can be explained by the resistive state of the SQUID that appears at $I^\sim > I_{c0}$, since the superconducting coil with FG, in the presence of AC, ceases to be a shunt with zero impendence, but becomes inductive resistance $x_L = 2\pi f L$ ($f = 75$ Hz $x_L \approx 0{,}5 \cdot 10^{-3}$ Ohm). As a result, the critical state of the SQUID is now determined by the summation of the half-period of AC and the magnitude of transport DC. With increasing AC, the fraction of DC in the formation of current steps decreases, thus creating steps of increasingly smaller length, until they disappear. This new state of the structure may be referred to as a postcritical quantum current state, as opposed to precritical $I^\sim < I_{c0}$. Clarification of the common factors of appearance and disappearance of current steps with increasing AC requires further studies.

Consider now the dependence $I_1(I_H)$ at different $I$ (Fig. 8). Figure 8 shows that the observed critical current $I_c$ of the SQUID decreases with increasing transport current $I$. Dependence of the critical current on $I$ is demonstrated in

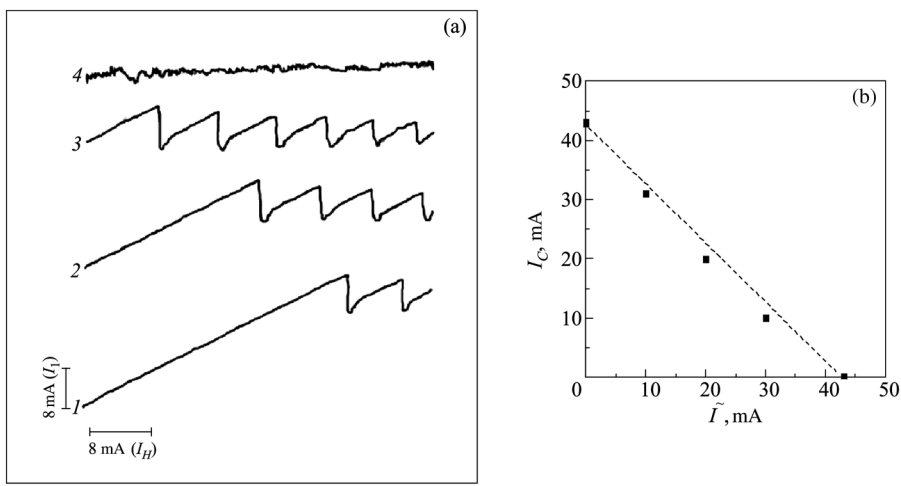

Fig. 9. Dependencies of current $I_1$ in inductance $L$ on current $I_H$ excited by the external magnetic field in the SQUID-SI structure circuit at different values of transport AC $I^\sim$, mA: 10 (*1*), 20 (*2*), 30 (*3*), 43 (*4*) (a); dependence of critical DC $I_c$ of the structure on transport AC $I^\sim$ (b).

Fig. 8(b). This type of dependence $I_1(I_H)$ at varying c0 is basically the confirmation of dependence at different values of $I_H$.

Therefore, the effect on the dependence $I_1(I_H)$ of current $I$ can be explained in common with the above explanation for the effect of field $H$ on the dependence $I_1(I)$. In this case, the dependence of critical current ($I_c$) of the SQUID on transport current $I$ is expressed as

$$I_c = I_{c0}(1 - I/I_{c0}). \qquad (7)$$

Finally, let us consider dependence $I_1(I_H)$ at various intensities of transport AC $I\tilde{\ }$ (Fig. 9).

Figure 9 indicates that the increase of AC $I\tilde{\ }$ may result in one of its half-waves adding up to DC ($I_H$) excited by the external magnetic field. This summation decreases the current $I_H$ required to achieve the critical state of the SQUID, therefore decreasing the external field $H$. It is thus clear that the observed effect $I\tilde{\ }$ on the critical state of the structure is similar to that of current $I\tilde{\ }$ on the above characteristic $I_1(I)$. The resulting dependence can thus be expressed as

$$I_c = I_{c0}(1 - I\tilde{\ }/I_{c0}). \qquad (8)$$

It should be noted that the effect of AC on all the described characteristics did not depend on its frequency in the range $f = 20$–$75$ Hz.

## 5. Conclusion

Measurement of current $I_1$ generated in the large inductance of the SQUID-SI structure with two different current flows passing through it significantly expanded the understanding of possible quantum states of this structure. In addition, these measurements made it possible to establish and explain the possibility of managing critical DC ($I_c$) of the SQUID in several ways. In particular, it was found that critical DC can be controlled using an alternating transport current ($I\tilde{\ }$) and alternating current ($I\tilde{\ }_H$) generated by the external alternating magnetic field. This feature makes it possible to predict the effect of real-life electromagnetic interference on the accuracy of measurement of the energy gap and relaxation time of the superconducting state by the interference method.

It has been found that the change in critical transport current $I_c$ of the SQUID and quantum 'steps' of current $I_1$ through inductance with the joint action of direct and alternating transport currents significantly depends on the ratio of $I\tilde{\ }$ and $I_{c0}$ ($I_{c0}$ is critical current in the structure at $I\tilde{\ } = 0$). At ($I\tilde{\ } + I$) < $I_{c0}$, AC only decreases the observed critical DC of the SQUID to zero, with the same size and periodicity of the quantum steps of current $I_1$. At $I\tilde{\ } > I_{c0}$, AC decreases the height and length of the steps to zero at $I\tilde{\ } \approx I_{c0} + \delta$ ($\delta$ is the step length at $I\tilde{\ } < I_{c0}$) and, accordingly, reduces their period. Thus, the existence of postcritical quantum states of the structure, in contrast to precritical quantum states, was established for the first time.

a)Email: bondarenko@ilt.kharkov.ua